\documentclass[twocolumn,pra,aps,superscriptaddress,showpacs]{revtex4}
\usepackage{amsmath}
\usepackage{graphicx}

\begin{document}
\title{Stable knots in the trapped Bose-Einstein condensates}
\author{Yong-Kai Liu}
\author{Shiping Feng}
\affiliation{Department of Physics, Beijing Normal University, Beijing 100875, China}
\author{Shi-Jie Yang\footnote{Corresponding author: yangshijie@tsinghua.org.cn}}
\affiliation{Department of Physics, Beijing Normal University, Beijing 100875, China}
\affiliation{State Key Laboratory of Theoretical Physics, Institute of Theoretical Physics, Chinese Academy of Sciences, Beijing 100190, China}
\begin{abstract}
The knot of spin texture is studied within the two-component Bose-Einstein condensates which are described by the nonlinear Gross-Pitaevskii equations. We start from the non-interacting equations including an axisymmetric harmonic trap to obtain an exact solution, which exhibits a non-trivial topological structure. The spin-texture is  a knot with an integral Hopf invariant. The stability of the knot is verified by numerically evolving the nonlinear Gross-Pitaevskii equations along imaginary time.
\end{abstract}
\maketitle
\section{introduction}
Topological objects are often interesting topics in various fields of physics ranging from the condensed matter physics\cite{Babaev,Yuki2} to the particle physics and the modern universe \cite{Juha,Vilenkin}. Although many works have been done, much less attention is paid to their mathematical existence\cite{ E1,E2,J.R1,J.R2,R.A.B,S.W,T.K,E.R}. In most works, the existence of the knot is discussed only qualitatively, using plausibility arguments, but the problem of constructing the corresponding solutions is rarely addressed. Among those, ultracold atoms provide an ideal pilot to study the complex topological excitations\cite{C.J}. In particular, two-component Bose-Einstein condensates (BECs) in which the interactions between the atoms can be precisely tuned by the magnetic-field Feshbach resonance, have been widely used to create topological defects\cite{S.B,S.T}.

The properties of the BECs are described by the order parameters within the mean field theory. By using the normalized spinor $\xi(\textbf{r})$ with $\xi^\dagger\xi=1$, the wavefunction is represented as $\psi(\textbf{r})=\sqrt{n(\textbf{r})}\xi(\textbf{r})$, where $n(\textbf{r})$ is the density of the condensate\cite{MH2,Yuki3,K.K}. When the symmetry group $G$ of a system reduces to its subgroup $H$ through spontaneous symmetry breaking, the topological excitations in the spinor condensates are characterized by homotopy classes of the order parameter (OP) space $M$ identified by the quotient space $G/H$\cite{K.K,MH2}. This examination can be carried out with the help of the homotopy groups of the OP space $\pi_n(M)$. The homotopy groups not only determine the topological invariants, but also stipulates the rules of coalescence and disintegration of the topological excitations\cite{Yuki4}.

The 3D topological structures are classified by the third homotopy group $\pi_3(M)$\cite{Yuki3}. Knots differ from other topological excitations such as vortices, skyrmions in that they are classified by a linking number while others are classified by winding numbers\cite{Yuki}. Knots are characterized by mapping from a three-dimensional sphere $S^3$ to $S^2$. The homotopy groups $\pi_3(S^2)=Z$.

In this paper, we construct an ansatz wavefunction which is the stationary solution in vanishing interactions limit. The state exhibits the topological structure with a knotted spin-texture. As the nonlinear interactions are switched on, we prove by numerical simulations that this topological structure is still energetically stable, providing conservation of particle number in each species atoms.

The paper is organized as follows. In Sec. II we present an exact solution in the limit of vanishing nonlinear interactions. In Sec.III we reveal the topological structure of stationary state. In Sec.IV we numerically verify the stability of the knot by taking into account the nonlinear coupling in the GPEs. A brief summary is included in Sec. V.

\section{stationary solution for non-interacting condensates}
We consider the two-component BECs that are confined in  a 3D trap. The dynamics is described by the coupled GPEs,
\begin{equation}\label{a}
i\hbar\frac{\partial\psi_i}{\partial t}=(-\frac{\hbar^2}{2m}\nabla^2+V(\textbf{r})+\sum_{j=1,2}U_{ij}\vert\psi_j\vert^2)\psi_i,
\end{equation}
where $\psi_i$ ($i=1,2$) denote the wave functions of the two components which are normalized to the number of atoms in each component, $N_1$ and $N_2$, respectively. The external potential is an axisymmetric harmonic oscillator $V(\textbf{r})=\frac{1}{2}m\omega^2(x^2+y^2)+\frac{1}{2}m\omega_z^2z^2$. The coupling constants $U_{ij}=4\pi\hbar^2a_{ij}/m$ represent the intra-species ($i=j$) and inter-species ($i\neq j$) interactions\cite{S.W}.

In order to simplify the equations, we adopt the dimensionless coupled GPEs\cite{S.W}.
By substituting $\psi_i(\textbf{r},t)=\psi_i(\textbf{r})\exp(-i\mu_i t) $, we obtain the stationary equations as
\begin{equation}\label{b}
\mu_i\psi_i=-\frac{1}{2}\nabla^2\psi_i+\tilde{V}(\textbf{r})+\sum_{j=1,2}\gamma_{ij}\vert\psi_j\vert^2\psi_i.
\end{equation}
We construct an ansatz wavefunction with a knot structure which is the stationary solution of Eq.(1) in vanishing interactions limit ($U_{ij}=0$), which is expressed in the cylindrical coordinates as,
\begin{equation}
\left(
          \begin{array}{c}
             \psi_1 \\
            \psi_{2}
          \end{array}
        \right)
=\sqrt{\frac{\omega_z}{\sqrt{\pi}}}e^{-\frac{1}{2}\omega_z^2z^2}e^{-\frac{1}{2}\omega^2\rho^2}
\left(
          \begin{array}{c}
           \sqrt{2}\omega_z z+i(1-\omega^2\rho^2) \\
           \omega\rho e^{i\phi}
          \end{array}
        \right).\label{ES1}
\end{equation}
By substituting it into Eq.(1) with $U_{ij}=0$, we can get $\omega_z=2\omega$, $\mu_1=4\omega$ and $\mu_2=3\omega$.

Figure.1(a) illustrates the density profile of the two components $\psi_1$ and $\psi_2$ for isosurfaces $n_{1,2}=0.02$, respectively. It shows that the $\psi_1$ component forms a density ring in the center surrounded which with a $2\pi$ phase changes (Fig.1(b) and (c)). It indicates a ring-like vortex whereas the component $\psi_2$ has a vortex line with its core along the $z$-axis.

\begin{figure}[h]
\includegraphics*[width=9cm]{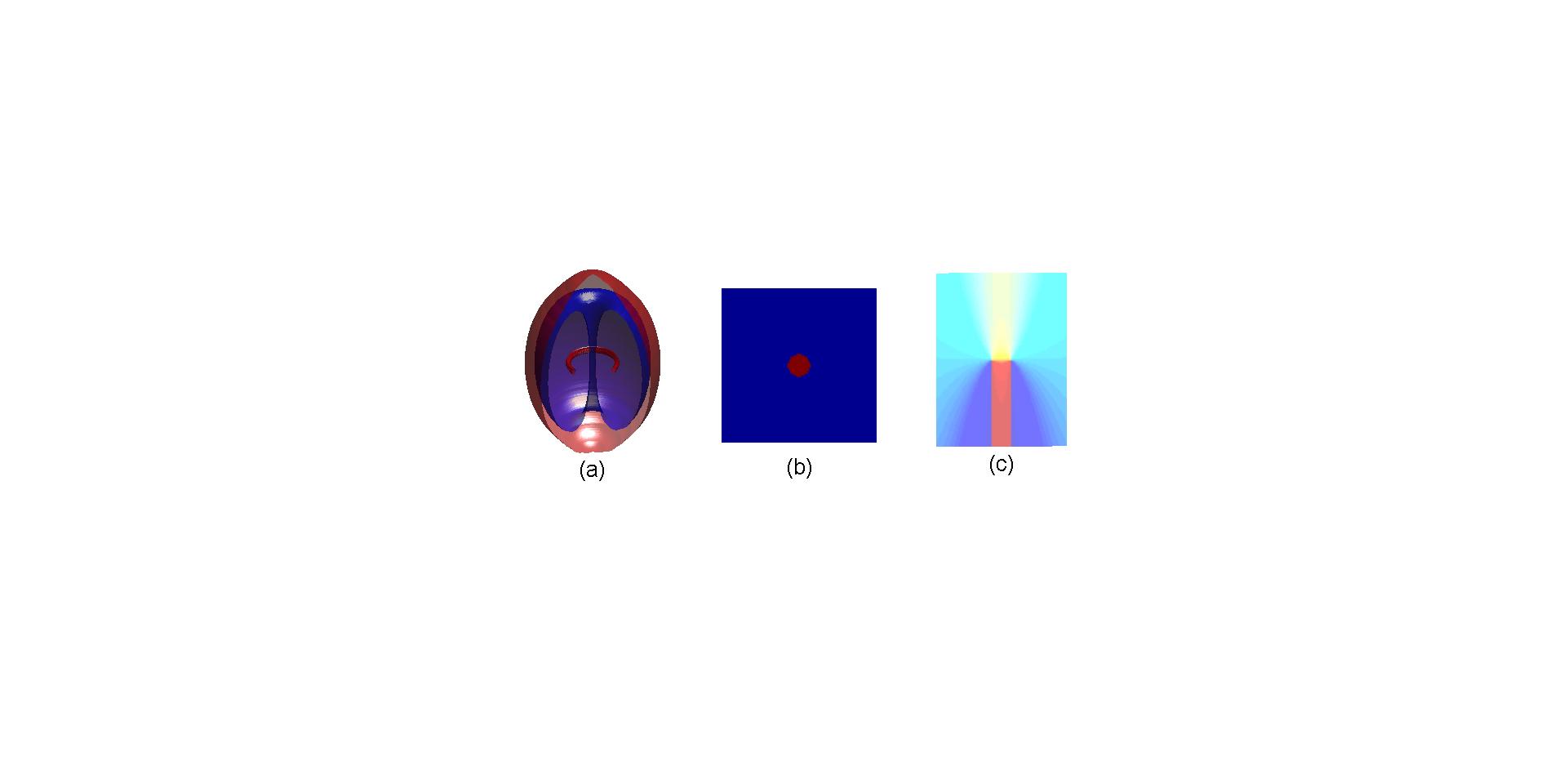}
\caption{(Color online) Density isosurfaces of the  $|\psi_1|^2$ (red) and the $|\psi_2|^2$ (blue) for $n=0.02$, respectively. The phase profiles of the $\psi_1$ in the $x-y$ plane (b) and in the $x-z$ plane (c) reveal a ring-like vortex in the $\psi_1$ condensate.}
\end{figure}

\section{topological structure}
In order to reveal the topological structure more clearly, we parameterized the order parameters of the condensates as\cite{Yang1}
\begin{eqnarray}
 \left(
          \begin{array}{c}
             \psi_1 \\
            \psi_{2}
          \end{array}
        \right)
=\sqrt{n(\textbf{r})}\left(
          \begin{array}{c}
           \cos(\theta(\textbf{r})/2)e^{i\varphi_1(\textbf{r})} \\
           \sin(\theta(\textbf{r})/2)e^{i\varphi_2(\textbf{r})}
          \end{array}
        \right),\label{Type G}
\end{eqnarray}
where $\theta\in[0,\pi]$, $\varphi_1\in[0,2\pi]$ and $\varphi_2\in[0,2\pi]$.
The local pseudo-spin $\textbf{S}$ which is defined by
$\textbf{S}=\xi^\dagger\boldsymbol{\sigma}\xi$, where $\boldsymbol{\sigma}$ is the Pauli matrix.

\begin{figure}
\includegraphics*[width=10cm]{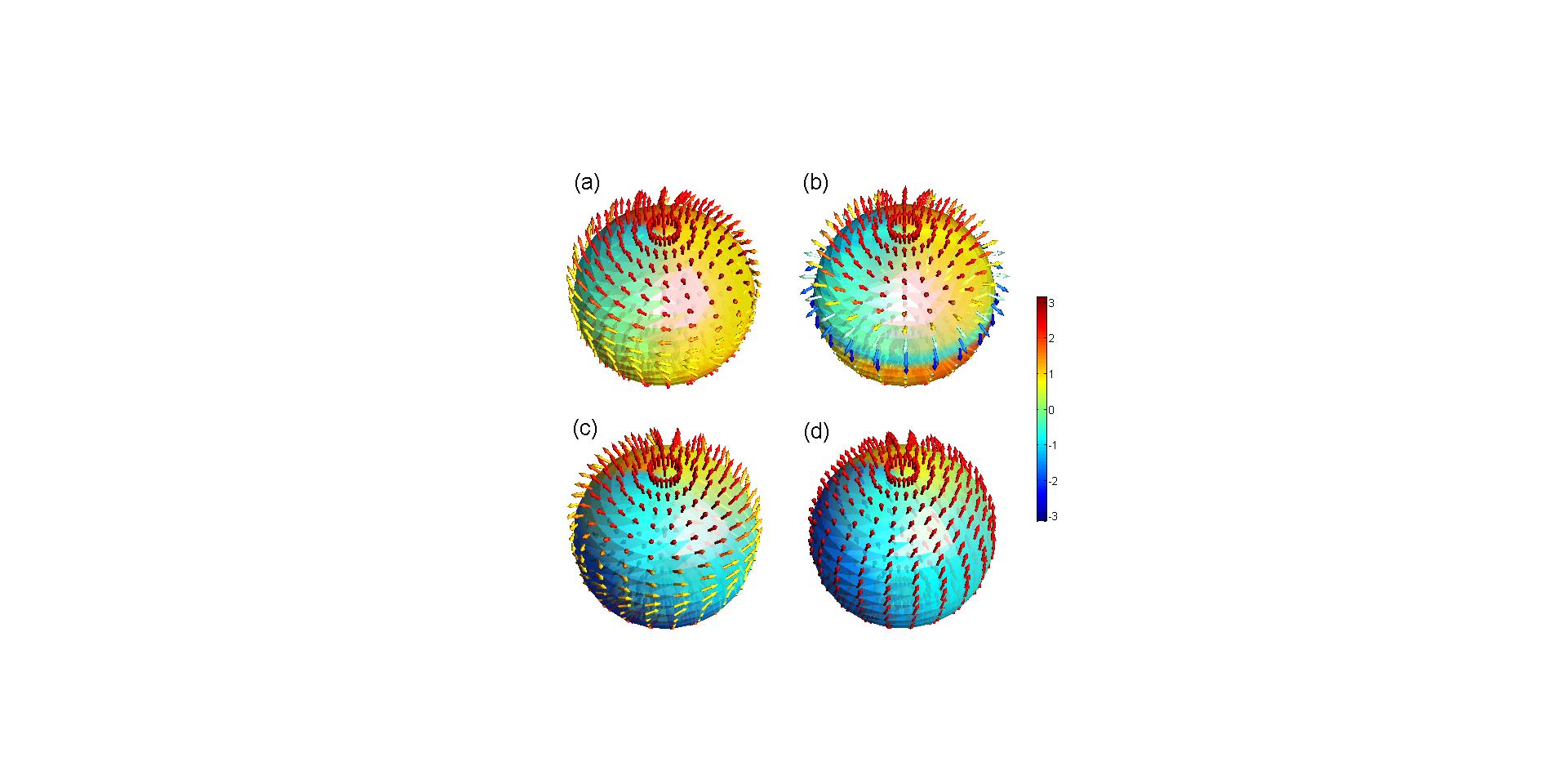}
\caption{(Color online) The pseudo-spin textures at different radii: (a) $r=1$, (b) $r=2$, (c) $r=4$, and (d) $r=12$. The color of the arrows indicate the values of $S_z$ ($\times \pi$). As radius is large enough, all spin point to the north.}
\end{figure}

The order parameter allows a topological classification $\pi_3(S^2)$ of a map from the real space to the spin vector which is described by
\begin{equation}
Q_\textrm{H}=\frac{1}{4\pi^2}\int \varepsilon_{ijk} \mathcal{F}_{ij} \mathcal{A}_k d^3 x,
\end{equation}
where the field strength $\mathcal{F}_{ij}=\partial_i\mathcal{A}_j-\partial_j\mathcal{A}_i={\bf S}\cdot(\partial_i{\bf S}\times\partial_j{\bf S})$\cite{Yuki2}. The integration is over the whole physical space where the density is assumed to be nonvanishing. By substituting the spin of Eq.(3) into the formula(5) of hopf charge, we can directly calculate $Q_H=1$.

We analyze the topological structure by using the cylindrically symmetric ansatz (\ref{Type G}) in comparison with the exact solution (\ref{ES1}). In the physical space, $\xi(\textbf{r})$ is given by the continuous deformation of the mapping $\theta(\textbf{r})=f_1(\varrho,z)$, $\varphi_1(\textbf{r})=f_2(\varrho,z)$, $\varphi_2(\textbf{r})=m\phi$.
For $f_1\in[0,\pi]$,$f_2\in[0,2n\pi]$, there is a winding number $n$ in the $\rho-z$ plane which is defined as $q=\frac{1}{4\pi}\int \epsilon_{ij} \sin\theta \nabla_i \theta \nabla_j \varphi_1  d^2\textbf{r}$. Thus, we can classify the topological excitations in terms of the pair of integers $(m,n)$.
 Figure 2 display the distributions of the spin fields at various radii $r=1,2,4,12$, respectively. The color of the arrows indicate the values of $S_z$ ($\times \pi$). As the radius tends to infinity, all the spins gradually point to the north direction, namely, $\textbf{S(\textbf{r})}\rightarrow (0,0,1)$, implying that the spin manifold is compactified to a point at spatial infinity.

\begin{figure}[t]
\begin{center}
\includegraphics*[width=9cm]{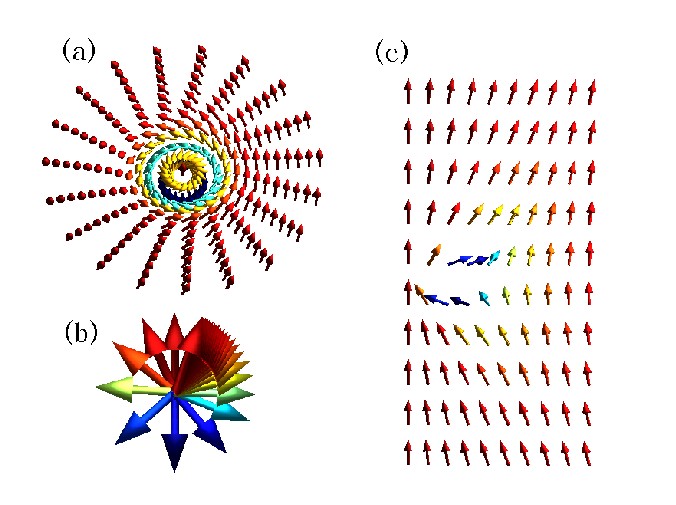}
\end{center}
\caption{(Color online) The spin-textures for (a) the horizontal section in $z=0$ plane and (b) along the radial direction. Each figure reveals a wind of the spin. (c) The spin-texture for the vertical section of $\phi=0$ (a half plane). It is a 2D skyrmion. The color on the arrows represent the value $S_z$.}
\end{figure}

In order to see the topology more clearly, the spin-textures in different sections are illustrated in Fig.3. Figure 3(a) shows the distributions of $\textbf{S(\textbf{r})}$ in the horizontal section of $z=0$. The spin rotates $2\pi$ around the azimuth direction ($\phi$). Figure 3(b) shows the rotation of $\textbf{S(\textbf{r})}$ along the $x$-direction. The spin twists $2\pi$ from $S_z=1$ in the center to $S_z=1$ at infinity, completing a whole winding in this direction. Figure 3(c) shows the distributions of $\textbf{S(\textbf{r})}$ in the vertical section of $\phi=0$. This section reveals a 2D skyrmion with winding number 1. The knot can be viewed as the 2D skyrmion($S^2$) rotating a circle($S^1$) around the $z$-axis. We can see the map as $S^3\rightarrow S^2\times S^1 \rightarrow S^2$, which can be thought as concentric spheres. As discussed above, we have a pair of integers (1,1) which indicates the topological invariant of the knot $Q_H=1\times 1$.
It can be verified by a deformation and recombination of Eq.(5) when substituting the  solution(3) into the formula of hopf charge. A similar analysis and deformation can be found in Ref.\cite{YM1}.

Apart from the direct calculation, hopf charge can also be viewed by the linking number from the image. As we know, $\textbf{S}$ defines a map from the 3D physical domain ($\textbf{x}$) into a 2D sphere. Consequently, the preimage of a point on the target $S^2$ corresponds to a closed loop in the compactfied $S^3$. The Hopf charge $Q_\textrm{H}$ characterize that the two loops corresponding to the preimages of any two distinct points on the target $S^2$ should be linked $Q_\textrm{H}$ times\cite{Yuki3}. Figure 4 illustrates the knotted features of the spin fields. The torus in Fig.4(a) is the isosurface of $S_z=0$. The red and green linked loops are the preimages of two points on the $S^2$ sphere. They are essentially two twisted tubes since we plot the isosurfaces of $S_x=0.98$ and $S_y=0.98$, respectively. In Fig.4(b), the torus is the isosurface of $S_z=0.98$, where the variations of color specify the angle of ($S_x,S_y$) which depicts the twist and chirality of the knot\cite{liu1}. The ring in the inner of the torus is the preimage of $S_x=0.98$. From the figures, we can see the linking number for this spin texture is 1, which indicate the hopf charge is one.

\begin{figure}[h]
\begin{center}
\includegraphics*[width=9cm]{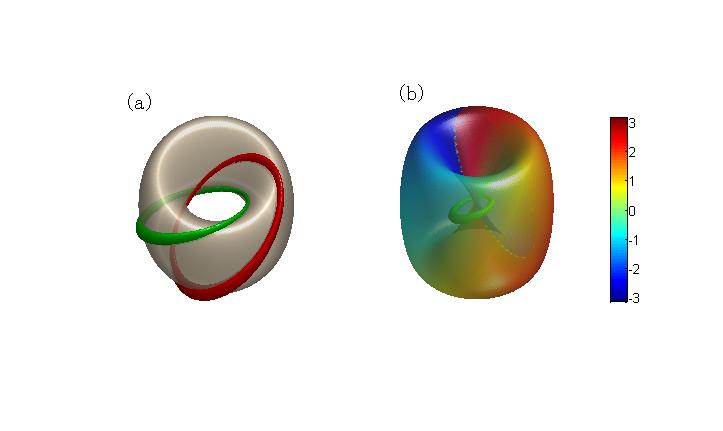}
\end{center}
\caption{(Color online) (a) The preimages of two points on the $S^2$ form two linked loops. The torus corresponds to the preimage of $S_z=0$. (b) The isosurfaces of $S_z=0.98$ (the torus) and $S_x=0.98$ (the inner ring). The variations of color on the torus represent the angle between $S_x$ and $S_y$ which reflect the twist and chirality of the knot.}
\end{figure}

\section{stability in the presence of nonlinear interactions}
In the previous works, a 'knot' or helical baby skyrmion whose physical space is $R^2\times S^1$ (periodic in z direction) is widely studied\cite{YM1,liu2}. But exactly, it is not a strict knot, as linking number or hopf invariant is not a homotopy invariant with the tube is not closed\cite{Juha,Nitta,E3}. Knot in BEC is rarely addressed, in the paper\cite{Yuki2}, a knot is created in the spin-1 BEC by manipulating an external magnetic field, however the lifetime for this knot is short.

Next, we numerically calculate the stability or dynamics of our knot by solving the time-dependent Gross-Pitaevskii equation. From the homotopy theory, the continuous deformation of Eq. (\ref{ES1}), with the satisfied boundary condition, does not change the topological charge.
In our case, by taking the immiscible regimes of interactions $a_{11}a_{22}< a^2_{12}$, and the number of atoms in each species separately conserved, we find the knot can be stabilized. The density ring in the $\psi_1$ component is fixed by the interaction of the two species, which prevent the knot from shrinkage.

The numerical simulation is carried out with the imaginary time evolution scheme\cite{E4,Bao1,Bao2,Bao3}. During each time step in the numerical simulation, we preserve the number of particle in the system while the chemical potential is adjusted correspondingly. This treatment has been widely used for studying various kinds of topological excitations in BEC system\cite{E4}.We use the split-step method with a spatial grid of $151\times151\times151$. The parameters of $^{87}Rb$, of which the scattering length can be tuned by the magnetic-field Feshbach resonance, are employed in the numerical calculations. We have tested different particle numbers with different proportion of $N_r=N_1/N_2$. For small number of atoms, the nonlinear repulsion between the two-species atoms are too weak to prevent the knot shrinking. So it is important to adjust the particle numbers and the interaction coefficients to keep the topological structure. Figure 5 shows the stable density distributions after a long-time evolution for $N_1=N_2=4.5*10^6$. The trap parameters are $\omega_z/\omega=2$ and $\omega=2\pi*7.8$ Hz. We note that the stability is insensitive to the shape of the trap. It holds for an isotropic harmonic trap.

\begin{figure}[h]
\begin{center}
\includegraphics*[width=9cm]{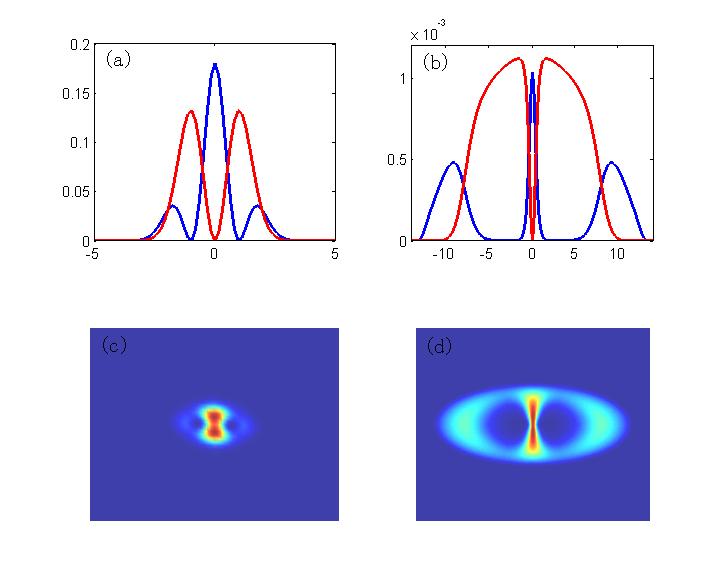}
\end{center}
\caption{(Color online) The density profiles of the exact solution (Eq.\ref{ES1}) for (a) $\psi_1$ (blue) and $\psi_2$ (red) along the $x$ axis and (c) $\psi_1$ in the 2D $x-z$ plane. (b) and (d) respectively are the corresponding density profiles of the two components after the imaginary time evolution of the initial state. The topological structure remain the same.}
\end{figure}

We have traced the topological invariants of  $Q_\textrm{H}$ during the imaginary time evolution and found it always kept unit for the stable knot. On the contrary, if we employ the total particle number conservation, the knot becomes unstable and will disappear. Proposals for experimental realizations of the knots can be found in Refs.\cite{Yuki2,Jy}.

\section{summary}
In summary, we have presented the exact 3D solution which exhibits a knot for the linear GPEs. The energetic stability of the topology is numerically demonstrated as the nonlinear interaction is switched on, provided the particle conservation of each component is adopted. The Hopf charge keep invariant during the dynamical evolution.

We thank W.Z. Bao for the helpful discussions on the numerical simulations. This work is supported by the NSF of China under grant No. 11374036 and the National 973 program under grant No. 2012CB821403.

\end{document}